\documentstyle[preprint,aps]{revtex}

\tighten

\begin{document}

\title{
Sedimentation and Flow Through Porous Media:
Simulating Dynamically Coupled Discrete and Continuum Phases
}

\author{Stefan Schwarzer\protect\footnote{e-mail: stefan@pmmh.espci.fr}}

\address{Laboratoire de Physique Mecanique des Milieux
  Heterogenes, Ecole Sup\'erieure de Physique et Chimie Industrielles,\\
  75231 Paris, Cedex 05, France}
\medskip
\address{H\"ochstleistungsrechenzentrum, Forschungszentrum
  J\"ulich,\\ 52425 J\"ulich, Germany}

\date{\today}

\maketitle

\begin{abstract}

  We describe a method to address efficiently problems of two-phase
  flow in the regime of low particle Reynolds number and negligible
  Brownian motion. One of the phases is an incompressible continuous
  fluid and the other a discrete particulate phase which we simulate
  by following the motion of single particles. Interactions
  between the phases are taken into account using locally defined drag
  forces. We apply our method to the problem of flow through random
  media at high porosity where we find good agreement to theoretical
  expectations for the functional dependence of the pressure drop on
  the solid volume fraction. We undertake further validations on
  systems undergoing gravity induced sedimentation.

\end{abstract}


\narrowtext

\section{Introduction}
\label{s:introduction}

A classical problem of chemical engineering is the understanding of
particulate two-phase flows, in which a continuous fluid constitutes
one component and a discrete particle phase the other. The practical
importance of understanding such a particle laden flow is evidenced by
its central role in geophysical phenomena like sand storms, dune
formation, or sediment transport, by the importance of biological
questions like the understanding of the flow properties of blood, or
cell component separation techniques as ultracentrifugation, by
industrial applications as diverse as pneumatic transport in tubes,
catalytic cracking, biological and chemical reactors, solid fuel
rocket motors, fluidized beds, sedimentation, filtration and many
more~\cite{b:Happel65,b:Soo67}.

If the discreteness of the particulate phase is fully taken into
account then the mathematical formulation of the flow problem involves
(i) a field equation for the continuous liquid phase --- in most cases
the Navier-Stokes equation, subject to a set of boundary conditions
both on the container walls and the surfaces of the suspended
particles --- and (ii) a set of differential equations for the time
evolution of the degrees of freedom of the individual constituents of
the particulate phase.  Due to the extremly complicated boundary
conditions and the nonlinearity of the underlying equations, an
analytical solution to this problem is impossible, except for
exceptionally simple cases. A numerical treatment, however, even with
simplifying assumptions, still poses tremendous practical problems.

Several simulation techniques have been developed which we briefly
review here.
(a) Finite volume techniques that implement no-slip boundary conditions
on the surface of each particle have been employed for very few
particles. The treated Reynolds numbers~\cite{Feng94} are in principle only
limited by the grid resolution and the computation time available.
Similarly, (b) lattice Boltzmann techniques for the liquid equation have
been used to simulate up to $1024$ suspended particles~\cite{Ladd94a}.
So far, these have come closest to provide realistic simulations
of two-phase flows with many particles. Most other approaches involve
certain assumptions that are only true in limited parameter ranges:
One important class of algorithms uses the Stokesian dynamics
technique~\cite{Brady88,Cichocki94} valid for low Reynolds number flow to
(c) obtain the flow field consistent with no-slip boundary conditions
\cite{Ladd88} or to (d) approximate the particles as being
point like~\cite{t:daCunha95,Ichiki93}.

(e) More popular in the engineering sciences, but less rigorous are
continuum approximations that involve two sets of
continuum equations, one for the liquid phase and one for the
particulate phase. The boundary conditions on the particle surface and
their influence on the flow is then represented by local drag forces
depending on the solid volume fraction and local velocity
differences~\cite{Jackson85,Batchelor88}. In these approaches there
are remaining open questions in the determination of the proper constitutive
equation for the solid and the momentum
exchange between the phases.

(f) To circumvent some of these later problems of the pure continuum
approaches, algorithms have come into fashion that combine a
discrete element~\cite{Cundall79} or molecular dynamics
description~\cite{b:Allen87} of the particles with a continuum
equation for the liquid as in (a), but use a drag term to couple the
two phases as in the pure continuum formulations
(e)~\cite{Tsuji92,Tsuji93}.

In this paper we will describe an algorithm of type (f). Such a method
allows immediate access to basically all physically relevant
quantities in the system, including particle coordinates and both
particle and liquid velocities, at computational costs comparable to a
``standard'' real space Wavier-Stokes integration. The main drawback
is that a neglect of the proper boundary conditions in the treatment
of the liquid will result in an inaccurate rendering of the short
scale flow properties. Since, however, our main focus will be the
ability of the algorithm to describe collective phenomena, i.e., the
effects that arise when the number of particles is large, we do not
have the ambition to describe accurately the local flow fields on the
scale of the particle size. At the moment the detailed effects of
neglecting certain aspects of the local flow are not clear to us.
Nevertheless we want to see which and how some collective phenomena
emerge directly from simple modeling assumptions --- as opposed to
using semi-empirical expressions as, e.g., done in
Refs.\cite{Tsuji92,Tsuji93}. In particular, we rely on the fact that
the long-range hydrodynamic interactions, which we presume to be the
most important for collective phenomena, are correctly represented by
the velocity and pressure fields according to the Navier-Stokes
integration under consideration of an additional field representing
the particle density.

The purpose of this paper is to give a detailed description of the
simulation algorithm that we use to model particulate two-phase flow
and its validation in the cases of (i) flow through a ``porous
medium'' consisting of the ``particles'' in the simulation which are
kept at fixed positions and (ii) of particles that undergo
sedimentation in a suspension under the influence of gravity.  In
Section \ref{s:model} we will first describe a two-dimensional
implementation of the algorithm. Then, in Sec. \ref{s:application},
we discuss first the case (i) of flow through a porous medium (Sec.
\ref{s:porous-media}) and second the case (ii) of sedimenting
particles (Sec. \ref{s:sedimentation}).

\section{The algorithm}
\label{s:model}

We aim at a simulation of macroscopic, hard sphere particles that
interact on contact when the interstitial fluid plays no important
role. In addition, they interact over long-range hydrodynamic forces,
mediated by an incompressible fluid medium between the particles. For
the moment, we consider a two-dimensional implementation of our
project. Since the analytic aspects of the solution of the
Navier-Stokes equation in two dimensions are very different from
the three dimensional solution, we expect at best a qualitative
agreement to the features of real three dimensional flows. However,
we feel justified to study the two-dimensional problem if it is
possible to find qualitative knowledge on that way.

We organize the description of our algorithm into three main parts.
In the first section we describe details of the
employed molecular dynamics technique for the particulate phase. The
second  compiles the fundamental equations
for the liquid phase and the details of their solution. In the third
section we elaborate on the particle-liquid
interaction.

\subsection{Molecular Dynamics of the Particle Phase}
\label{s:model:md}

\subsubsection{Single particle properties and forces}

Each particle $i$ in the simulation is characterized by its radius
$r_i.$ We take the particle size distribution to be slightly
polydisperse with the radii $r_i$ drawn from a Gaussian distribution
$h_p(r/\bar{r})$ with mean $\bar r$, which is cut off at its standard
deviation $p\bar{r}$,
\begin{equation}
\label{eq:distribution}
 h_p(r/\bar{r}) \propto \left\{
 \begin{array}{ll}
             \exp \left[-\frac{1}{2}
             \frac{(r/\bar{r}-1)^2}{p^2}\right],
                             & \mbox{ if}\quad |r/\bar{r} -1| < p, \\
             0,      & \mbox{ else.}
  \end{array}
  \right.
\end{equation}
Here, the distribution is written in terms of the dimensionless radius
$r/\bar{r}$ and the polydispersity parameter $p.$

We disregard effects of rotational motion of the particles
and only consider their
translation. Thus we describe the particle motion only by the
coordinates of their  center of mass $\vec{x}_i = (x_i, y_i).$
Although the particle centers are constrained to lie in the $xy$
plane, we will assume --- for reasons that we address in Sec.
\ref{s:model:coupling} --- that the particles can otherwise be
regarded as three dimensional.  Therefore we associate with each
particle a mass $m_i = (4/3) \pi r_i^3 \rho_p$, where $\rho_p$ is the
constant particle density.

We consider the following forces to be acting on particle $i$,
\begin{equation}
  \label{e:ftot}
  \vec{F}_i = \frac{4}{3} \pi r_i^3 (\rho_{\rm p} - \rho_{\rm l}) \vec{g}
    + \vec{F}^w_i + \vec{F}^d_i +
    \sum_j (\vec{F}^n_{ij} + \vec{F}^t_{ij}).
\end{equation}
Here, $\vec{F}_i$ is the sum of all forces on $i$, the
$\vec{F}_i^w$ are forces due to the presence of the boundary, the term
proportional to $\vec{g}$ accounts for particle weight and buoyancy. The drag
force $\vec{F}_i^d$ will be discussed in Sec. \ref{s:model:coupling}. The
sum runs over the other particles in the system, but is effectively
restricted to the neighborhood of particle $i$, since we will only
introduce short range interparticle forces in radial ($\vec{F}^n_{ij}$) and
and tangential ($\vec{F}^t_{ij}$) direction in the next two sections.

Container walls, if present, manifest themselves by elastic forces and
frictional forces in analogy to those acting between two particles.
We will use for them the equations that result from those for two
collision partners --- to be introduced below --- in the limit of the
other particle having infinite mass and radius.

In direction of $\vec{g}$, we account for the weight of the particles.
Acting opposite to gravity, we add the buoyancy reaction of the
liquid, which equals the weight of the displaced fluid of constant
density $\rho_l$. Our coordinate system is chosen such that $\vec{g}$
points in $-y$ direction.

The forces from Eq. (\ref{e:ftot}) enter the equations of motion for
the particulate phase, $m_i \ddot{\vec{x}}_i = \vec{F}_i.$ We solve
this coupled system employing a fourth order predictor-corrector
algorithm as described, e.g., in~\cite{b:Allen87}.

\subsubsection{Pairwise interparticle forces in radial direction}

At very low Reynolds numbers the lubrication forces between particles
play an important role. Since they have a divergent character when the
particles come close, particles in principle do not touch in this
regime. However, if the fluid in the system is a gas or the particle
based Reynolds numbers are not any longer small or if the particles
are assumed to have some surface roughness, then they will touch and
we have to model the contact forces between them.

If no fluid were present then the particles are force free unless
they touch, whereupon strong repulsive and dissipative forces act
between them, resulting from the viscoelastic properties of the
particles. To model these forces, we here use contact models
frequently employed in granular matter
research~\cite{Cundall79,Walton92,Ristow95}. The first of the forces
acting on member $i$ of a particle pair $i,\;j$ in contact is an
elastic restoring force $\vec{F}^{\rm el}_{ij}.$ This force is
proportional to the {\it virtual overlap\/} $\xi_{ij} \equiv
(r_i + r_j) - |\vec{x}_i-\vec{x}_j|$
of the particles. If the overlap
is positive, then
\begin{equation}
  \label{e:f_el}
  \vec{F}^{\rm el}_{ij} = -k_n \xi_{ij}
     \vec{n}_{ij},\quad\mbox{for~} \xi_{ij} > 0.
\end{equation}
Here, $\vec{n}_{ij} = (\vec{x}_j - \vec{x}_i)/|\vec{x}_j - \vec{x}_i|$
denotes the unit vector pointing from the center of particle $i$ to
the center of particle $j$ and $k_n$ is the stiffness of the contact,
which we assume to be constant.\footnote{
  The determination of $k$ for realistic contacts in 3D is not a
  trivial matter. The value of $k$ depends on the exact physical
  processes governing the contact and thus in general both on material
  constants as the Young modulus $E_{\rm p}$ or the Poisson number
  $\sigma_{\rm p}$ and the geometry of the particles. Many researchers use
  the nonlinear Hertzian contact theory, in which $\vec{F}^{\rm el}_{ij}
  = - k_n \xi_{ij}^{3/2} \vec{n}_{ij}$. For equal sized spheres, the Hertzian
  theory gives $k_n= \sqrt{2r} E_p/ 3(1-\sigma_p^2).$ We refer the
  reader to the tables \ref{t:flow} and \ref{t:sed} for values used in
  the simulations.
}
As mentioned above the force vanishes if the particles do
not overlap, i.e., $\xi_{ij} < 0.$

To take the dissipative character of the contact into account, we add
a velocity dependent friction term to the elastic restoring force.
This damping force shall also act in the direction of the line
connecting the particle centers (the {\it normal\/} direction) and be
proportional to the normal relative particle velocity. On inclusion of
this term, we obtain for the contact force $\vec{F}^{c,n}_{ij}$ in
normal direction,
\begin{equation}
  \label{e:fn1}
  \vec{F}^{c,n}_{ij} =
  (-k_n \xi_{ij} - \gamma_n m_{\rm red}
    (\dot{\vec{x}}_j - \dot{\vec{x}}_i)\cdot \vec{n}_{ij}) \vec{n}_{ij}.
\end{equation}
In this relation, $m_{\rm red}$ is the reduced mass $m_im_j/(m_i+m_j)$
of the pair in contact and $\gamma_n$ determines the strength of
the dissipation. We have suppressed the indices on $m_{\rm red}$ and
$\gamma_n,$ but we consider them to vary among different particle pairs.

Equation (\ref{e:fn1}) is the equation for a damped harmonic
oscillator while the particles are in contact.  For a given initial
normal relative particle velocity $\vec{v}^n_i$ it can be solved
analytically for the velocity after the contact $\vec{v}^n_f.$ Since
energy is dissipated, the ratio of these velocities is less than one;
its value is termed {\it restitution coefficient\/} $e \equiv |
\vec{v}^n_f|/ |\vec{v}^n_i| $. One obtains for a specific particle
pair (pair indices suppressed),
\begin{equation}
  \label{e:restitution}
  e =
  \exp \left[ - \pi / \sqrt{\frac{4}{\gamma_n^2}
                                \frac{k_n}{m_{\rm red}}-1}
       \right].
\end{equation}

However, in the presence of a liquid at low Reynolds numbers, the
dominant force at small distances between pairs of particles is the
lubrication force $\vec{F}^{\rm l}$ arising from the pressure
necessary to replace the liquid within the gap between the two
approaching spheres. The lubrication force damps the relative motion
of the particles very strongly and diverges when the particles touch.
Since in our approximate treatment of the liquid flow in particular
the lubrication force is not well represented we add it as an
additional component of the particle-particle interactions, only
active at short distances between the particles.
The normal component of the lubrication force is~\cite{Davis95}
\begin{equation}
  \label{e:lubrication1}
  \vec{F}^{\rm l,n}_{ij} = - 6 \pi \eta \frac{r_{\rm
      red}^2}{(-\xi_{ij})} [(\dot{\vec{x}}_i - \dot{\vec{x}}_j)\cdot
  \vec{n}_{ij}] \vec{n}_{ij}.
\end{equation}
In the above equation, $\eta$ denotes the shear viscosity of the liquid
and $r_{\rm red}$ the reduced radius $r_i r_j/(r_i + r_j)$ of the
pair. Again we have suppressed the indices $i$ and $j$ for simplicity.
The expression (\ref{e:lubrication1}) is only valid for small positive
separations between the surfaces of the two involved particles.  These
are reflected in negative overlaps $\xi_{ij}$ of small modulus.

To take into account the limited range of validity of Eq.
(\ref{e:lubrication1}), we cut off $\vec{F}^{\rm l,n}_{ij}$ where
$(-\xi_{ij}) > r^{\rm red}.$ To avoid a discontinuity in the
force law, we subtract in Eq. (\ref{e:lubrication1}) a constant equal
to the value of the force just at this cutoff distance.

Furthermore, we remove the divergence at $\xi_{ij} = 0$ by adding to
the value of $(-\xi_{ij})$ in the denominator a small positive number
$\delta r^{\rm red}$. We have used a value of $\delta = 0.1.$ Other
than just being a numerical contrivance our physical motivation is
found in the unavoidable surface roughness of particles in reality,
which may cause the particle to come into contact despite the
lubrication force.

Similarly, contacts due to numerical inaccuracies are,
almost unavoidable in a dense system with many particles. To cover
these spurious cases as graceful as possible, we have matched the
$\gamma_n$ of Eq. (\ref{e:fn1}) such that the force law is
continuous when $\xi = 0,$ for particles of radius $\bar{r}.$
We obtain,
\begin{equation}
  \label{e:gamman}
  \gamma_n = 6 \pi \eta \frac{r_{\rm red}}{m_{\rm red}}
                \left(\frac{1}{\delta} -
                  \frac{1}{1+\delta}\right) .
\end{equation}

The verbal statements in the preceding three paragraphs
condense into the following equations for the total interparticle
forces in the normal direction, comprising both contact and
lubrication forces,

\begin{eqnarray}
  \label{e:interparticle}
  \vec{F}^n_{ij} &=& \vec{F}^{c,n}_{ij} + \vec{F}^{l,n}_{ij} \nonumber \\
   & = & \left\{
  \begin{array}{ll}
    \vec{0}, & \mbox{if~} (-\xi_{ij}) > r_{\rm red}, \\{}%
      [ - 6 \pi \eta r^2_{\rm red} ( \dot{\vec{x}}_j -
       \dot{\vec{x}}_i ) \cdot \vec{n}_{ij}]
       [\frac{1}{-\xi_{ij} + \delta r_{\rm red}}
        - \frac{1}{(1+\delta)r_{\rm red}}] \vec{n}_{ij},
        & \mbox{if~} 0 < (-\xi_{ij}) < r_{\rm red}, \\{}%
    [-k_n \xi_{ij}  - \gamma_n m_{\rm red}
 (\dot{\vec{x}}_j - \dot{\vec{x}}_i)\cdot \vec{n}_{ij})] \vec{n}_{ij},
             &\mbox{if~} \xi_{ij} > 0.
  \end{array}
  \right.
\end{eqnarray}

This force is continuous over the whole range of
$\xi_{ij}.$

\subsubsection{Pairwise interparticle forces in tangential direction}

To model the frictional forces acting perpendicular to the line
connecting the two particle centers --- the {\it tangential\/} forces
--- we resort, as in the case of the normal forces, to notions of
particle contact modeling. Here, the Coulomb law of sliding friction
asserts that the magnitude of the tangential friction force
$\vec{F}^t_{ij}$ is --- on contact --- proportional to the magnitude
of the acting normal force $\vec{F}^n_{ij},$ [from Eq.
(\ref{e:interparticle})] with a constant of
proportionality $\mu$ usually between $0.05$ and $0.5$,
\begin{equation}
  \label{e:coulomb}
  |\vec{F}^t_{ij}| = \mu |\vec{F}^n_{ij}|.
\end{equation}
This force is always directed opposite to the relative motion.
Numerical problems may occur in near central impact, when the
tangential component of the relative velocity is small, but
$\vec{F}^n$ is large. Then the likewise large tangential component of
the force resulting from (\ref{e:coulomb}) may cause an unphysical
oscillatory behavior of the tangential velocity during contact. We
therefore replace Eq.  (\ref{e:coulomb}) for small relative tangential
velocities by a velocity proportional friction term. Thus, finally,
the tangential friction force on particle $i$ becomes
\begin{equation}
  \label{e:ft}
  \vec{F}^t_{ij} = - \min(\mu |\vec{F}^n_i|,
                       \gamma_t |\vec{v}^t_{ij}|)
                        \frac{ \vec{v}^t_{ij}}{|\vec{v}^t_{ij}|},
\end{equation}
where $\vec{v}_{ij}^t$ has been introduced as an abbreviation for the
relative tangential velocity $(\dot{\vec{x}}_i - \dot{\vec{x}}_j) -
[(\dot{\vec{x}}_i - \dot{\vec{x}}_j)\cdot\vec{n}_{ij}]\vec{n}_{ij}$.
We do not include shear contributions of the lubrication force into
the interparticle forces. For simplicity, $\gamma_t$ is taken to be
constant.

\subsection{Fluid Model}
\label{s:model:liquid}

We describe the state of the fluid phase by three continuum fields,
namely (i) the velocity field $\vec{u}(\vec{x})$ of the fluid, (ii)
its pressure $p(\vec{x})$ and (iii) a field $\epsilon (\vec{x})$ equal
to the local volume fraction of liquid. These variables have only
physical meaning as averages over volume on a scale larger than
that of the individual particles. Their choice is motivated by continuum
approaches to multiphase flow \cite{Jackson85}.

The position and geometry of the particles determines the field
$\epsilon (\vec{x})$ which --- for specific discrete tiling of the
simulation plane --- is a more or less smooth function varying between
$1$ (no particles) and $0$ (full occupation by particles), defined for
all tiles and all times. Similarly, the time evolution of the velocity
field $\vec{u}(\vec{x})$ is determined by the pressure distribution,
viscous contributions and a force distribution $\vec{f}(\vec{x})$
which comprises both volume forces on the liquid and momentum exchange
contributions with the particulate phase (Sec.~\ref{s:model:coupling}).

We follow Ref.~\cite{Jackson85} and write for the time evolution of
the liquid velocity $\vec{u}$ (we will drop the argument $\vec{x}$ of
the fields from now on),
\begin{equation}
  \epsilon \rho_{\rm l}\left[ \frac{\partial\vec{u}}{\partial t} +
  (\vec{u}\,\nabla)\vec{u}\right] =
  -\epsilon \nabla p + \epsilon\eta\,\nabla^2\vec{u}
                                    + \epsilon \vec{f}.
  \label{e:nse}
\end{equation}
Although $\epsilon$ drops out from this equation, it enters into the
the momentum exchange contribution to $\vec{f},$ in the sense that the
momentum transfer to the particulate phase due to the fluid phase ---
due to drag between particles and liquid --- must be ``fed back'' to a
liquid volume smaller than in the case without particles.

Since we have in mind applications to systems with typical velocities
much smaller than the velocity of sound, we can assume that the fluid
phase is incompressible. The equation of liquid mass continuity then reads
\begin{equation}
  \frac{\partial \epsilon \rho_l}{\partial t} + \nabla \cdot (
  \epsilon \vec{u}) = 0.
  \label{e:ce}
\end{equation}
Equation~(\ref{e:ce}) presents a constraint on the velocity field that
must be fulfilled at all times and may be employed to obtain the
pressure field via an iterative procedure which we model after the
artificial compressibility method of
Chorin~\cite{Chorin67,Chorin68,b:Peyret83}. Here, we sketch the basic
ideas of its 2D implementation briefly and refer the reader for more
details to the literature.

We discretize the differential equations (\ref{e:nse}) and
(\ref{e:ce}) in the following way. We place the velocity components
$u_x,\; u_y$ as well as the pressure $p$ on three quadrilateral meshes
with lattice spacing $\Delta x.$ With respect to the pressure grid,
the grids for the $x$ and $y$ velocity components are shifted by
$\Delta x/2$ in $x$ and $y$ direction respectively. This construction
is commonly refered to as the MAC mesh and has several computational
advantages. For instance, it is a simple means to avoid numerical
instabilities due to mesh decoupling~\cite{b:Peyret83}. The choice of
location for the computational quantities is conceptually related to
location of variables in finite volume techniques for flux
conservative differential equations, in which the fluxes are located
on the corresponding faces of a control volume whereas the conserved
quantities themselves reside in the center of the
volume~\cite{b:Patankar80}.

We obtain the pressure and the velocity components by an iterative
procedure. Let the index $n$ refer to values at time $t=t_{n}$ and
$n+1$ to those at $t=t_{n+1}=t_n + \Delta t$ after a timestep of
duration $\Delta t$. The index $k$ shall denote an iteration index. We
define $p_{n+1,0} \equiv p_{n},$ i.e., we start an iteration for the
new pressure at time $t_{n+1}$ with the old values at $t_{n}.$ We
obtain a tentative velocity field at $t=t_{n+1}$ from an evaluation of
the discretized Navier-Stokes equation~(\ref{e:nse}),
\begin{equation}
 \rho_{\rm l}\frac{\vec{u}_{n+1,k+1} - \vec{u}_n}{\Delta t}  =
      -\rho_{\rm l}(\vec{u}_n \nabla)\vec{u}_n -
            \nabla p_{n+1,k} + \vec{f}_n + \eta\nabla^2\vec{u}_n.
  \label{e:v-iter-initial}
\end{equation}
where the symbol $\nabla$ now denotes second order precise
difference operators on the
lattice.  As mentioned above, the solid volume fraction enters
implicitly through $\vec{f}_n.$
However, since (\ref{e:v-iter-initial}) is a discretized Navier-Stokes
equation, its stability criteria on the MAC grid have been studied and
are reported, e.g., in \cite{b:Peyret83,Kopetsch89}. These
criteria state that the values $\Delta x$ and $\Delta t$ are subject
to the two constraints
\begin{equation}
 \Delta t \le \frac{4\eta}{\rho_{\rm l}(|u_x^{\rm max}| + |u_y^{\rm max}|)^2},
\end{equation}
and
\begin{equation}
   \Delta t \le \frac{\rho_{\rm l}(\Delta x)^2}{4\eta}.
\end{equation}

In general, the velocity field $\vec{u}_{n+1,k+1}$ resulting
from~(\ref{e:v-iter-initial}) considered together with $\epsilon_n$
does not satisfy the continuity equation~(\ref{e:ce}). Rather, one has
to conceive the continuity equation as a constraint that determines
the pressure field within the fluid such that the resulting velocity
field satisfies the continuity equation at all times.

To this end, one derives an iterative procedure to determine an
appropriate pressure field. This procedure is based on the idea that
the local violation of the continuity equation, i.e., the value of the
left hand side of Eq. (\ref{e:ce}), can be used to correct the
pressure field. The correction is taken in a direction such that the
modulus of the violation is reduced in the next iteration step, after
a new tentative velocity field has been determined.  We write
\begin{equation} \label{e:p1}
  p_{n+1,k+1} = p_{n+1,k} -
    \lambda\rho_{\rm l}[\frac{\partial \epsilon_n}{\partial t}
    + \nabla\cdot(\epsilon_n\vec{u}_{n+1,k+1})].
\end{equation}
Here, a large value of the parameter $\lambda$ is crucial for rapid
convergence. The value of $\lambda$ is, however, by stability
requirements constrained to
\begin{equation}
   0 < \lambda \le \frac{(\Delta x)^2}{4\Delta t}.
\end{equation}
In the simulations we have chosen $\lambda$ to equal its upper
stability limit.

It should be noted that the values $\epsilon_n$ in Eq. (\ref{e:p1})
are not all located on the same subgrid: the time derivative is taken
at the pressure points and the $\epsilon_n$ multiplying
$\vec{u}_{n+1,k+1}$ is realized by different fields, each living on
the same subgrid as the associated velocity component.

Once we have determined a new pressure field, we need to recalculate
new velocity values consistent with the $p_{n+1,k+1}.$ These
velocities may be obtained using the Navier-Stokes equation
(\ref{e:v-iter-initial}), or equivalently, avoiding the costly
reevaluation of Eq. (\ref{e:v-iter-initial}), using the relation
\begin{equation} \label{e:p2}
   \rho_{\rm l}\frac{\vec{u}_{n+1,k+2} - \vec{u}_{n+1,k+1}}{\Delta t} =
       \nabla ( p_{n+1,k+1} - p_{n+1,k} ).
\end{equation}

Iterating Eqs.~(\ref{e:p1}) and~(\ref{e:p2}) yields a new pressure
field and after convergence a velocity field for time $t_{n+1}$ such
that the equation of continuity is satisfied.

For our purposes, the described algorithm has three advantages. It (i)
generalizes straightforwardly to 3D and (ii) unlike spectral or
streamfunction methods it gives immediate access to the quantities $p$
and $\vec{u}$ in real space. Fast access to the latter is crucial for
the calculation of the particle-liquid interaction. Moreover (iii),
only the chosen coarseness of the spatial and temporal discretization
limits the range of Reynolds numbers addressable in the simulation.
However, since the presence of the particles is communicated to the
liquid among others through the field $\epsilon$, it does not make
much physical sense to use a computational grid on a scale smaller
than the particle size.

For Reynolds numbers requiring such smaller grids, one could think
maybe of an approach to decompose the liquid equation into an equation
for the average flow and additional equations describing the
fluctuations around it, in the spirit of turbulence modeling
\cite{b:Launder72}.

\subsection{Interaction of Particles and Fluid}
\label{s:model:coupling}

The main problem in simulations of multiphase flow that try to bypass
the specification of boundary conditions on the phase boundaries is to
specify an expression for the momentum exchange between fluid and
particulate phase. Even for a single sphere this task is
daunting~\cite{Maxey83}. For extended fixed random collections of
spheres the phenomenological Ergun formula~\cite{Ergun49,Ergun52}
gives the pressure gradient as a function of solid fraction and liquid
velocity. Tsuji and coworkers \cite{Tsuji92,Tsuji93} have used a
``localized'' form of the Ergun equation to estimate the force of the
liquid on a sphere and obtain for pneumatic transport in pipes and
fluidized beds qualitative agreement of flow patterns and quantitative
agreement of some quantities.

Having the complicated situation in mind, and being aware of the
significant approximation involved, we wish to explore the
consequences of a very simple model for the momentum exchange. We use
a local version of the ``global'' Stokes formula --- which is just one
term in the expression given by Maxey~\cite{Maxey83} for the drag on
an isolated sphere under low Reynolds number conditions ---
to obtain the drag force acting on a sphere of radius $r_i$,
\begin{equation}
  \label{e:stokes}
  \vec{F}^d_i = - 6\pi \eta r_i [\dot{\vec{x}}_i - \vec{u}(\vec{x}_i)].
\end{equation}
In order to evaluate the liquid velocity field at the location of the
particle we interpolate linearly the velocity values from the closest
four grid points. That is to say, if the pair $(x,y)$ denotes the
coordinates of one of these four ``adjacent'' grid points $\vec{x}_i$
then $w(x,y) \equiv (1-|x_i-x|/\Delta x)(1-|y_i-y|/\Delta x)$ is the
weight associated with $(x,y)$; the index $i$ refers to the particle
location.  We obtain thus for some velocity component $u$ the
interpolated value $u(\vec{x}_i)$ as $u(\vec{x}_i) = \sum
w(x,y) u(x,y)$. The sum extends over the four corners of the MAC grid
plaquette associated with the component $u$ into which $\vec{x}_i$
falls.

As equation (\ref{e:stokes}) resembles the 3D expression for the drag
on an isolated {\it sphere}, we should give here a motivation for this
modeling assumption.  A ``pure'' 2D simulation should, strictly
speaking, be one of rigid, parallel {\it cylinders}. However, at a
fixed small Reynolds number, the drag per unit length of a cylinder
\cite{b:Landau89} does not depend on its radius. This behavior is very
different from the observations in 3D experiments, where there is a
strong dependence of the drag on the size of the spheres.  Thus, in
particular, effects of the particle polydispersity cannot be expected
to be represented properly by a model based on drag-forces of
cylinders.

We therefore use Eq. (\ref{e:stokes}) for the drag and refer the force
(\ref{e:stokes}) to a {\it reference length} $z$ equal to the
average particle diameter. The picture we have in mind behind this
procedure is that we imagine the particle configuration repeating itself
all $z$ length units, and the liquid being infinitely extended in $z$
direction, however both particles and liquid being constrained to
motion in the $xy$ plane.  when we calculate the drag per unit volume
that enters into the Navier-Stokes equation.  The finite drag force
(\ref{e:stokes}) acts on the liquid at the location of the center of
the particle and we must thus ---
in order to convert it to the force density required in the continuum
equation --- represent it by a $\delta$-function
at $\vec{x}_i$,
\begin{equation}
  \vec{F}^d_i\frac{1}{z\epsilon(\vec{x})} \delta (\vec{x}-\vec{x}_i),
\end{equation}
in the Navier-Stokes equation. The $\epsilon(\vec{x})$ ensures that
the force density is refereed to the liquid fraction alone which is
necessary to conserve momentum. We form the sum over all particles
and add the uniform contribution of gravitation to
obtain the full volume force density term,
\begin{equation}
  \label{e:volume_force}
  \vec{f}(\vec{x}) = \vec{g} +
     \sum_i \vec{F}^d_i\frac{1}{z\epsilon(\vec{x})} \delta (\vec{x}-\vec{x}_i).
\end{equation}

On our computational lattice we implement the expression
(\ref{e:volume_force}) by distributing $\vec{F}^d_i/z (\Delta x)^2
\epsilon (\vec{x})$ to the four grid points closest to $\vec{x}_i$.
To this end, we employ the same weights as introduced for the
interpolation of the velocity components above. For example, the
contribution to point $(x,y)$ is $(1-|x_i-x|/\Delta
x)(1-|y_i-y|/\Delta x) \vec{F}^d_i/z (\Delta x)^2 \epsilon (\vec{x}).$

This concludes the description of the implementation of our algorithm.

\section{Application to example problems}
\label{s:application}

We will now describe the results of the application of the described
algorithm on selected physical problems in order to validate our
approach, and assess its limitations or respectively the
consequences of using the simple drag law (\ref{e:stokes}).

\subsection{Flow through porous media}
\label{s:porous-media}

As our first system, we consider the motion of a fluid through a {\it
  fixed\/} random assembly of particles. Our mental picture is that
the particle assembly is a model for a random porous medium with very
high porosity. In the simulation, we then have to keep the particles
pinned to their initial positions. However, we calculate all
forces acting between particles and liquid, but we only update the
liquid's degrees of freedom in the computational timestep. Excluded
volume effects due the presence of the particle phase and local
frictional drag still influence the fluid motion.
Gravity is set to zero, or equivalently the flow is considered to
lie in the horizontal plane.

\begin{figure}[h]
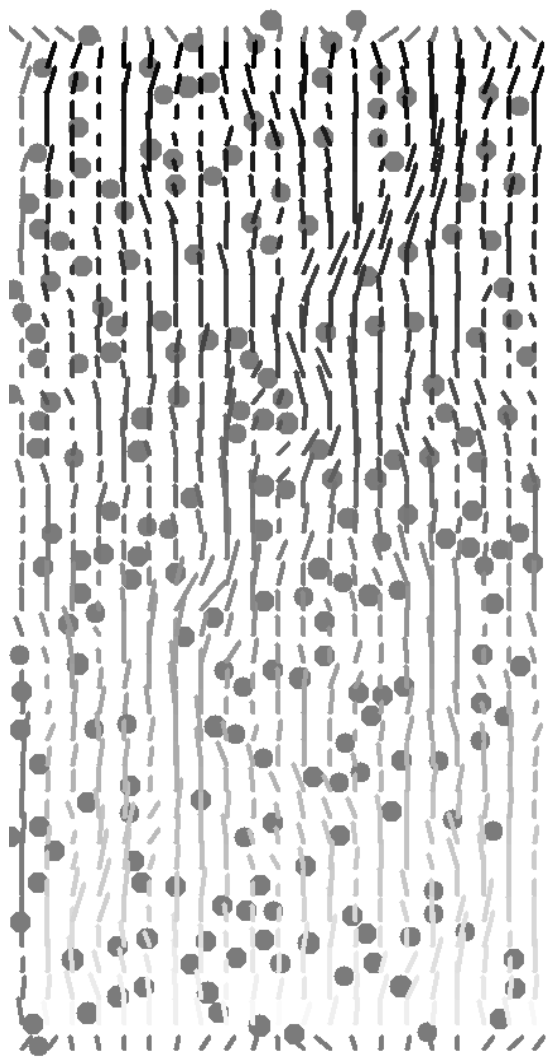

\caption{
\label{f:porous-pic}
2D Fluid flow pattern through a random assembly of ``spheres.''  In
this figure the 2D area fraction is $\bar{\epsilon} = 0.15$ and the
particle based Reynolds number $Re_{\rm p} = 0.01.$ Lines indicate the
direction and the magnitude of the flow.  Clearly visible are the
effects of the mass conservation on the flow: The flow concentrates in
regions with few particles and ``engulfs'' the particles on smaller
scales.}
\end{figure}

In $x$ direction the system has the width $L_x$ and boundary
conditions are periodic. We impose fixed superficial flow velocities
$u$ in $y$-direction at the inlet and outlet of the system at $y=0$
and $y=L_y$, where $L_y$ denotes the height of the system. A typical
arrangement of particles in a small system of $L_x / \bar{r} \approx
33$ and $L_y / L_x = 2$ at $Re_{\rm p} = 3.75 \times 10^{-3}$ and the
resulting stationary flow pattern is displayed in Fig.
\ref{f:porous-pic}. The circles indicate the particles, the arrows
direction and magnitude of the liquid flux obtained by multiplying the
local liquid volume fraction with the flow velocity. We see how the
particle volume fraction influences the flow pattern such that the
current concentrates in regions with few particles present.  Note that
the fluid velocity is defined everywhere in space, even at points
covered by particles and that the flow velocities should therefore be
considered as average values over the specific grid cell.  Note also
that the flux vectors are not displayed at their location used for
computational purposes but are extrapolated to and displayed together
with the color coded pressure at the location of the pressure points.

\begin{figure}[h]
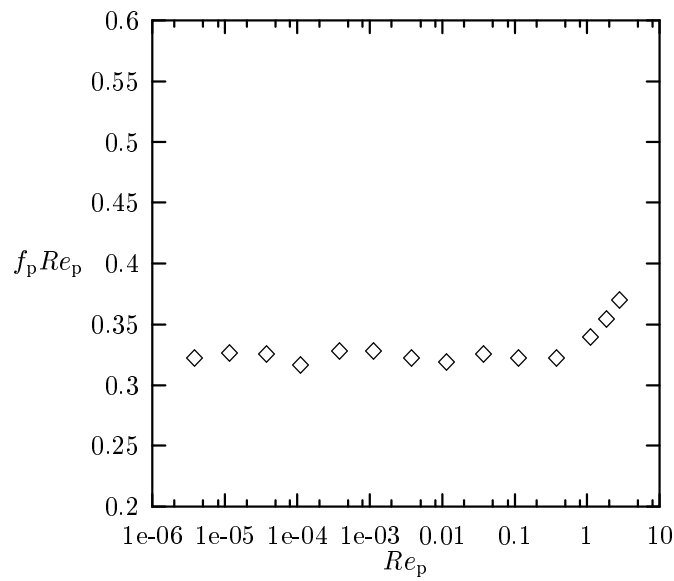

\caption{
\label{f:porous-darcy}
We plot the product $f_{\rm p} Re_{\rm p}$ of friction factor and
particle Reynolds number vs. the particle Reynolds number $Re_{\rm
  p},$ at constant porosity $\bar{\epsilon} = 0.93$~ $(= 1-\sum_i (4/3)
\pi r_i^3 / L_x L_y z).$ Apart from small statistical fluctuations due
to different initial placements of the particles, the curve is
constant in the regime of Reynolds numbers $\ll 1$, thus showing that
the pressure drop is proportional to $u$ as required by Darcy's law.
}
\end{figure}

In this system, we measure the overall pressure drop per length
$\Delta p/L_y$ as a function of the overall liquid volume fraction
$\bar{\epsilon}$ of the medium and the superficial liquid velocity
$u$.  In the viscous regime the pressure drop is proportional to the
fluid velocity $u$.  We evidence a constant ratio of pressure drop to
fluid velocity in Fig.~\ref{f:porous-darcy} for several orders of
magnitude of the particle Reynolds number $Re_{p} \equiv \rho_{\rm l}
\bar{r} u/ \eta$ at fixed ``porosity'' $\bar{\epsilon}$.

\begin{figure}[h]
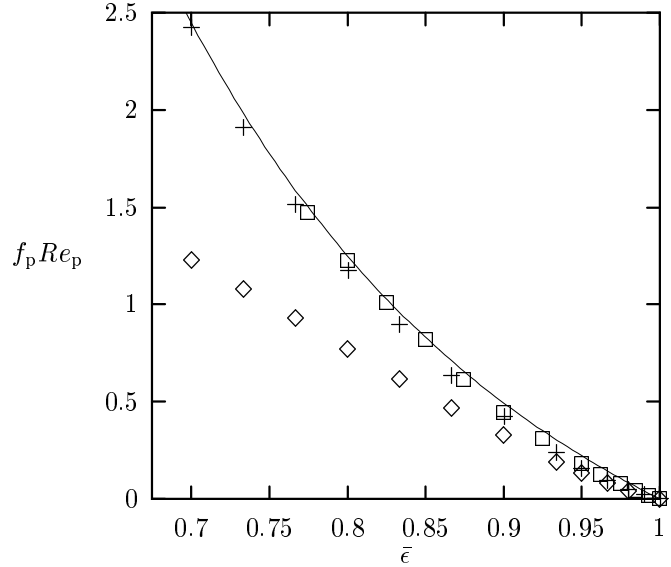


\caption{
  \label{f:porous-pdrop}
  We plot the product $f_{\rm p} Re_{\rm p}$ of friction factor and
  particle Reynolds number vs. the volume fraction $\bar{\epsilon}$.
  Different curves correspond to different flow rates (Reynolds
  numbers) and to different values of the system ``depth''
  $z/\bar{r}=2~(+),$ $z/\bar{r}=8/3~(\Box).$ A third set of runs has been
  made for the case of only frictional coupling between fluid and
  obstacles, setting the liquid volume fraction to one everywhere
  $(\Diamond)$.  The volume fraction $\bar{\epsilon}$ is calculated as
  3D volume fraction, considering the particles as spheres and the
  simulation box as quadrilateral of extension $L_x \times L_y \times
  z$.}
\end{figure}

We report our results in terms of the friction factor
\begin{equation}
  \label{e:friction-factor}
  f_{\rm p} = -\bar{r} \Delta P / \rho_{\rm l} u^2 L_y,
\end{equation}
which is, due to the linearity of the pressure drop in the viscous
regime, inversely proportional to the particle Reynolds number
$Re_{\rm p} \equiv \rho_{\rm l} \bar{r} u/ \eta$.  Accordingly we have
plotted in Fig. \ref{f:porous-pdrop} the product $f_{\rm p} Re_{\rm p}$
which --- in the viscous regime --- is independent of the Reynolds
number. The product's value is plotted against the liquid volume
fraction or porosity $\bar{\epsilon}$ based on the 3D volume of the
particles divided by the box volume $L_x \times L_y \times z$, where
$z$ denotes the effective box depth introduced
in Sec. \ref{s:model:coupling}. The three different curves in
the plot indicated by symbols correspond to two different values of $z/\bar{r}
= 2$ and $8/3$ and one run (lowest lying curve) where only drag
between particles and fluid was considered in the simulation and the
local liquid fraction $\epsilon (\vec{x})$ in Eqs. (\ref{e:nse}) and
(\ref{e:ce}) was kept equal to $1.$ The curves for different $z/\bar{r}$
collapse into a single universal curve. We consider this data collapse
as an {\it a posteriori\/} justification of the picture of
the simulated system which we have used to find a form for the
momentum feedback to the liquid (Sec. \ref{s:model:coupling}).

Concerning the functional form of the result in 3D, the literature
lists several phenomenological expressions for the porosity dependence
of the friction factor~\cite{b:Dullien79}. Popular expressions for
$f_{\rm p} Re_{\rm p}$ include (i) the half-empirical Carman-Kozeny
relation $f_{\rm p} Re_{\rm p} \sim (1-\bar{\epsilon})^2/\bar{\epsilon}^3$ and
(ii) the phenomenological Rumpf-Gupte form $f_{\rm p} Re_{\rm p} \sim
\bar{\epsilon}^{-5.5}.$ Both relations are only valid in an intermediate
range of porosity and do not apply to the limit of very high porosity
or very low solid volume fraction, where one expects $f_{\rm p}
Re_{\rm p} \sim 1-\bar{\epsilon}$ to leading order for the following
reasons.  For a strongly ``dilute'' random medium, let us consider the
solid as independent spherical ``defects.'' Each defect exerts the
Stokes force $6\pi r \eta \tilde u$ on the fluid, where $\tilde u$ is the
interstitial fluid velocity. Thus we approximate, on the one hand, the
internal rate of energy dissipation to be $6\pi r \eta \tilde u^2 N,$
where we denote the total number of 'defects' in the system by $N.$ On
the other hand, externally, the pressure drop across the system and
the superficial fluid velocity $u = \tilde u \bar{\epsilon}$ give the rate at
which work is done on the system to be $\Delta P L_x z u$. Equating the
two rates and using $(1-\bar{\epsilon}) = (4/3) \pi r^3 N / (L_x L_y z)$
yields
\begin{equation}
  \label{e:stokes-drop}
 f_{\rm p} Re_{\rm p} = \frac{9}{2} \frac{1-\bar{\epsilon}}{\bar{\epsilon}^2}.
\end{equation}
This equation correctly predicts the pressure drop to vanish in the
limit of large porosity, but it is clear from the argumentation given
above that its validity is restricted to the regime of large liquid
volume fraction.

The solid line in Fig. \ref{f:porous-pdrop} shows the prediction of
formula (\ref{e:stokes-drop}). The simulation data compares very
favorable with this prediction for the whole range of simulated
effective 3D porosities $\bar{\epsilon} > 0.7.$ Please note that
(\ref{e:stokes-drop}) has no adjustable parameters. The agreement is
surprising considering the rather crude assumptions that we have made.

In contrast, the Carman-Kozeny formula predicts the pressure drop to
vanish quadratically as $\bar{\epsilon} \to 1$, and the Rumpf-Gupte formula
predicts even a constant pressure drop for $\bar{\epsilon} \to 1$.  However,
both formulas are applicable only in the intermediate or low porosity
regime so that the lack of agreement with (\ref{e:stokes-drop}) is not
surprising.

\subsection{Sedimentation}
\label{s:sedimentation}

\begin{figure}[h]
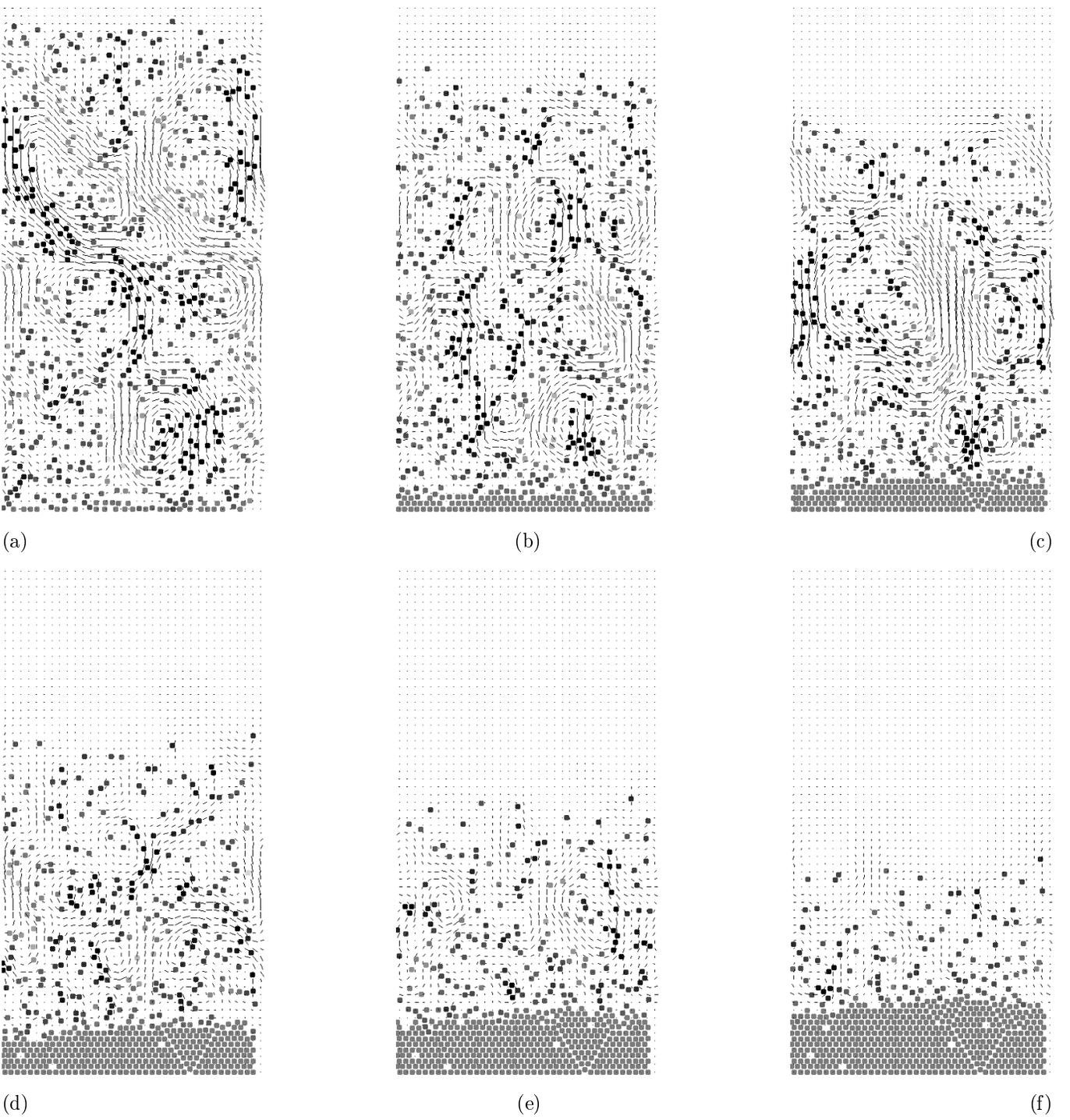

%
\caption{
\label{f:sed-conf}
Typical particle configurations during a batch settling simulation at
$t V_S / \bar{r} = 5.2$ (a), $16$ (b), $26$ (c), $36$ (d), $46$ (e),
and $56$ (f) where
$t$ denotes time, $V_S$ the Stokes velocity and $\bar{r}$ the average
particle radius.  The system size is chosen to be comparatively small
$L_x/\bar{r} = 80$ and $L_y/L_x = 2$; 606 particles are visible and
the 2D solid area fraction $\sum_i \pi r_i^2 / L_x L_y$ is $0.10$.
Lines indicate the direction and amplitude of the fluid flow.
%
%
Shades of gray denote the $y$-component of the particle velocity, dark
particles move down and light ones up. It is interesting to see that
there is substantial internal motion of particles in complicated
vortex patterns. The particle settling is visible ``on average'' for
example at the upper sedimentation front.}
\end{figure}

We now discuss the application of our algorithm to the more general
case when both particles and liquid are mobile and important dynamical
consequences arise from the coupling of the two phases.  A prototype
example for this case is batch settling sedimentation.  Imagine a
homogeneous mixture of particles and liquid being placed into a
quadrilateral --- or for our purposes rectangular --- container and
being initially at rest. If there exists a density difference between
particles and liquid, then gravitational driving forces will set the
particles in motion. Complicated fluid-mediated hydrodynamic
interactions between the particles lead to convoluted trajectories and
give rise to collective phenomena as for example anisotropic self
diffusion of particles.  The particles slowly settle to the bottom of
the container with an average speed $\langle V (\Phi) \rangle$ that
decreases as the volume fraction $\Phi$ of particles in the container
increases. We have chosen here $\Phi = 1-\bar{\epsilon}$ to follow
the convention in most of the literature on sedimentation.
We display a typical situation during the batch settling
process in Fig.  \ref{f:sed-conf}.

The conditions of the simulation are periodic or respectively no-slip
boundary conditions in the horizontal $x$ direction and no-slip
boundary conditions in vertical $y$-direction. Gravity is taken to act
in $-y$ direction. One denotes as the hindered settling function $f_{\rm
  hs}(\Phi)$ the ratio of the sedimentation velocity $\langle V
(\Phi) \rangle$ to the Stokes velocity $V_S$,
\begin{equation}
  \label{e:def-hs}
  \langle V \rangle / V_S \equiv f_{\rm hs} (\Phi ),
\end{equation}
where
\begin{equation}
  \label{e:stokes-velocity}
  V_S = (2/9) (\rho_{\rm p} - \rho_{\rm l}) g \bar{r}^2 / \eta,
\end{equation}
which is the settling velocity of an isolated sphere in
an infinitely extended fluid.

We investigate $f_{\rm hs} (\Phi)$ for an almost monodisperse system.
The particle Reynolds number $Re_{\rm p} = \rho_{\rm l}
\bar{r} V_S / \eta$ is smaller than $1$, typically
$\approx 3\times 10^{-2}$ and (ii) the ``simulation box'' Reynolds number
$Re_{\rm b} = \rho_{\rm l} L_x V_S / \eta$ is $100$ times larger,
i. e. $Re_{\rm b} \approx 3 $.
Furthermore, we choose the particle size large enough such
that the effects of Brownian motion may be neglected, corresponding to
the regime of high (iii) Peclet number.
We have performed simulations and determined the settling velocity as
a function of the solid fraction of the particle suspension. The
following four different sets of conditions for
``thought'' experiments have been performed to assess the
role and importance of lubrication and backflow for the simulations.
In particular, simulation series

\begin{itemize}
\item[(i)] includes lubrication effects [Eq. (\ref{e:lubrication1})], effects
  of particle void fraction [Eqs. (\ref{e:nse},\ref{e:ce})], and
  periodic boundary conditions in $x$ direction, perpendicular to
  gravity;

\item[(ii)] differs from series (i) only in the respect that we have not
  considered the lubrication term
  (\ref{e:lubrication1}). Without these lubrication forces, we choose the
  value of $\gamma_n$ in Eq.~(\ref{e:fn1}) such that for pair
  collisions a restitution coefficient of $0.9$ results [cf. Eq.
  (\ref{e:restitution})];

\item[(iii)] differs from (ii) in the respect that we have additionally
  set the liquid fraction $\epsilon(\vec{x})$ to $1$, as if the
  particles consisted of liquid and not of a separate solid phase.
  The interaction of particle and liquid phase results only from the
  pointlike frictional drag between the two.

\item[(iv)] differs from (i) only in the respect that we have used
  no-slip boundary conditions at $x=0$ and $x=L_x.$
\end{itemize}

\begin{figure}[h]
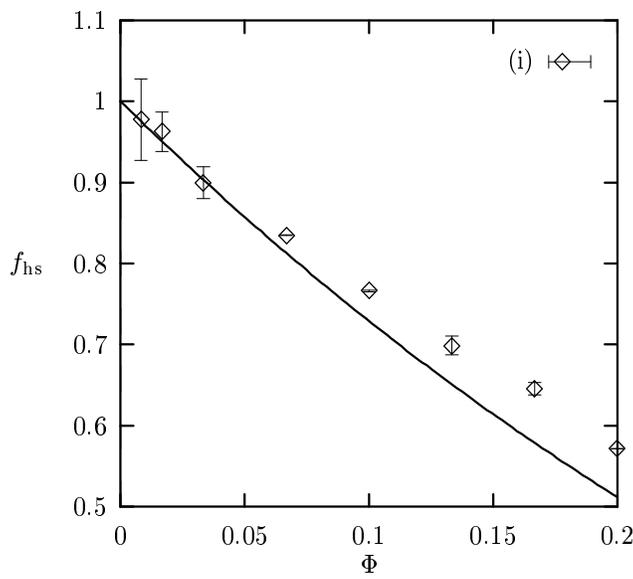

\caption{
  \protect\label{f:sed-vel} Hindered settling function $f_{\rm hs}
  \equiv \langle V\rangle / V_S$ for simulation series (i) (see text),
  including lubrication, void fraction and using periodic boundary
  conditions in $x$ direction. The ordinate shows the effective 3D
  solid fraction $\Phi$, which is related to the 2D area fraction of
  particles in a simple way, given that the radius distribution of the
  particles is monodisperse: $\Phi = (4/3) (\bar{r}/z) \Phi_{\rm 2D}.$
  The data points have been averaged over three runs with different
  initial conditions.
  The solid line is the form $f_{\rm hs} = (1-\Phi)^{3}$
  proposed in the text.}
\end{figure}

Figure \ref{f:sed-vel} displays the hindered settling function of
case (i). The ``Stokes'' velocity used to normalize the data has been
obtained from simulations of single spheres. In addition, we have
plotted a theoretical expression (\ref{e:hindered_settling}), $f_{\rm
  hs}(\Phi)= (1-\Phi)^3$, which we will discuss later in the text
(Sec. \ref{s:-problems}).  It
should be noted that experiments in the viscous regime --- both
$Re_{\rm b}$ and $Re_{\rm p}$ much smaller than $1$ --- often report a
correlation with a $(1-\Phi)^n,~n\approx 5$ dependence
\cite{Nicolai95,Xue92}, the Richardson-Zaki formula
\cite{Richardson54}. It may be that in our case the larger value of
$\Re_{\rm b}$ prevents us from seeing $n\approx 5$.  Of course, also
the nature of hydrodynamic interactions in 2D is quite different from
the 3D case, such that probably only full 3D simulations see $n=5$
\cite{Ladd94}.

\begin{figure}[h]
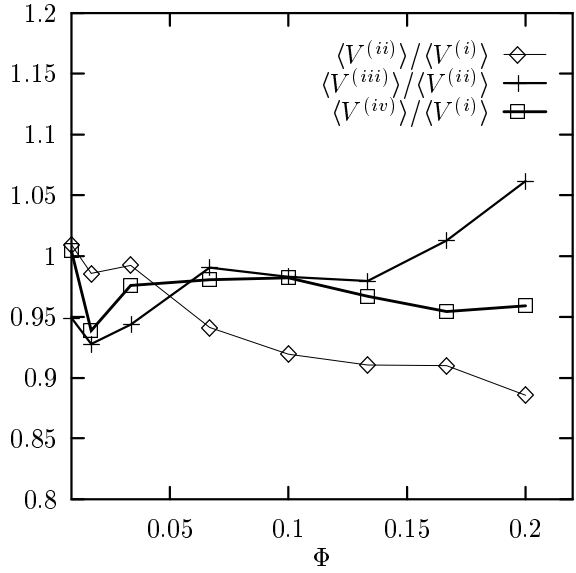

\caption{
  \protect\label{f:sed-rat} Ratios of sedimentation velocities vs. the
  effective 3D volume fraction $\Phi$ in the simulation computed under
  four different sets of simulation conditions (i)..(iv), for details
  see text. Diamonds ($\diamond$) denotes the ratio the computed
  velocity disregarding lubrication (ii) to the velocity in the
  ``full'' simulation (i). Similarly, ($+$) denotes the ratio of
  series (iii) to (ii) --- no lubrication and in (iii) additionally
  the liquid fraction set to $1$.  Boxes ($\Box$) denote the ratio
  (iv) to (i) for the difference between periodic and no-slip boundary
  conditions for the box walls.  All data points have been computed
  using velocity averages over three runs with different initial
  conditions.
}
\end{figure}

The graphs in Figure \ref{f:sed-rat} address the question in which way
the resulting sedimentation velocities in series (ii) to (iv) differ
from that in (i). To this end, we have calculated the ratios of the
sedimentation velocities. First, the symbol ($\Diamond$) denotes the
ratio $\langle V^{(ii)}\rangle / \langle V^{(i)}\rangle $.  Since
lubrication alters the interactions between particles more
significantly when the particle fraction is high, we see that the
fraction deviates more and more from $1$ as the solid fraction
increases. The ratio becomes smaller than $1$ which means that the
system ``without'' lubrication sediments slower.  We understand this
behavior taking into account the strongly damping character of the
lubrication force.  Such Damping forces between particles have the
effect to reduce the {\it relative\/} velocity between them and thus
to favor the creation of loosely connected particle agglomerates.
These effects are well known for ``dry'' granular systems with
dissipative contact interactions between particles
collisions~\cite{Goldhirsch93,McNamara93}. These particle clusters
then trap some liquid within and fall as almost coherent units. Such a
unit displays a larger Stokes velocity than its constituent particles
since $V_S$ is proportional to the squared radius of the falling
object.

Second, we assess the effect of backflow in the simulation by setting
$\epsilon (\vec{x}) =1$ in the Navier-Stokes and the continuity
equation (\ref{e:nse},\ref{e:ce}). The particles are then effectively
point-like as far as the fluid is concerned. Denoted by $\Box$, the ratio
$\langle V^{(iii)}\rangle / \langle V^{(ii)}\rangle $ starts at a
value below one --- which results from particularities in the momentum
exchange modeling (see Sec. \ref{s:-problems}) --- and increases with
volume fraction. We can estimate the effect of backflow by assuming
that the average relative velocity of particles and liquid phase
are the same in situations (ii) and (iii): A short calculation
assuming that the average relative velocity between particles and
liquid is the same in both cases
predicts $\langle V^{(iii)}\rangle / \langle V^{(ii)}\rangle
= 1/(1-\Phi) \approx 1 + \Phi.$
The observed increase is less steep as expected from this relation
indicating a more complicated effect as consequence of the introduction of
the volume fraction field, which we do not understand at this point.

Third, we have performed a simulation with no-slip boundary conditions
on the container walls instead of the periodic boundary conditions
used otherwise. We denote the data points for $\langle
V^{(iv)}\rangle / \langle V^{(i)}\rangle $ by the symbol
$\Diamond$. The ratio is, apart from a point at very low volume
fraction, smaller than one. The liquid velocity is constrained to be
zero  at the container walls and hence we believe that altogether the
motion of the liquid is less vehement than in the periodic case.
Since particle and liquid motion are very strongly coupled --- the
distance that a particle has to travel in order to reach a terminal
velocity is much less than a particle diameter --- the on average
smaller particle velocities at the container walls suffice to
slow down the settling.

\subsubsection{Relation of sedimentation and flow through a random medium}

At this point it is interesting to note that the functional forms
proposed both for the volume fraction dependence of
the mean settling velocity in the sedimentation problem and for the
pressure drop in the problem of flow through random media (the
Rumpf-Gupte relation) are probably not independent from each other.

More precisely, it is possible to relate the formulas for the pressure
drop in flow through a random medium and the settling velocity in the
sedimentation problem. During the sedimentation process, one observes
two shock fronts in the suspension. One front separates the densely
packed particle assembly at the bottom of the vessel from the bulk of
the suspension, the other the bulk from the leftover clear fluid
region at the top. If the particles were frozen in their instantaneous
positions by compensating the forces acting on them, then the pressure
drop between these two shock fronts is just the hydrostatic pressure
drop of the pure liquid. In an infinitely extended system, time
averaged, each particle experiences a total force equal to the
difference of its weight and buoyancy. If the particles move freely
then an additional pressure difference between the two shock fronts
arises because now the liquid must support the particles.  Therefore,
in steady state, the arising pressure difference must just equal the
difference of particle weight and buoyancy, i.e.,
\begin{equation}
\label{e:dp1}
  \frac{\Delta P}{L} = \Phi ( \rho_{\rm s} - \rho_{\rm l} ) g.
\end{equation}
Here $L$ denotes the height difference within the two mentioned shock
fronts over which the pressure drop is measured.

On the other hand, if the particles were immobile, we could apply a
drag formula of the form introduced in Sec. \ref{s:porous-media} for
flow through random media, say $f_{\rm p} Re_{\rm p} = f_{\rm
  rm}(\bar{\epsilon}),$ to find the pressure drop $\Delta P/L.$ Let in this
case $\tilde u$ denote the average relative velocity of liquid and
particles. Then we find that
\begin{eqnarray}
\label{e:dp2}
\frac{\Delta P}{L} & = &
          \frac{\tilde u \eta}{\bar{r}^2} f_{\rm p} Re_{\rm p}
          \nonumber \\
          & = & \frac{\tilde u \eta}{\bar{r}^2} f_{\rm rm}(\bar{\epsilon}).
\end{eqnarray}
There is no net flux of material (liquid plus particles) through an
arbitrary horizontal cross section of the suspension. Thus, if the
particles settle, a net upward flux of fluid with velocity $-\langle V
\rangle \Phi / (1-\Phi)$ results. The average relative velocity of
particles to liquid is therefore $\tilde u=\langle V \rangle /
(1-\Phi).$ We now equate the expressions (\ref{e:dp1}) and
(\ref{e:dp2}), substitute $\tilde u,$ and solve for $\langle V
\rangle$:
\begin{equation}
\label{e:average_v}
   \langle V\rangle = \bar{r}^2 (\rho_{\rm s} - \rho_{\rm l})
   \frac{g}{\eta} \frac{\Phi (1-\Phi)}{f_{\rm rm}(1-\Phi)}.
\end{equation}
Here, we have replaced the liquid volume fraction $\bar{\epsilon}$ by
the solid volume fraction $\Phi = 1-\bar{\epsilon}.$ We obtain the
dimensionless hindered settling function $f_{\rm hs}(\Phi)$ by
division of this equation by $V_S$,
\begin{equation}
  \label{e:result-relation}
  f_{\rm hs}(\Phi) \equiv \frac{\langle V\rangle}{V_S} =
    \frac{9}{2} \frac{\Phi (1-\Phi)}{f_{\rm rm}(1-\Phi)}.
\end{equation}

For example, if we insert the functional form of $f_{\rm
  rm}(\bar{\epsilon})$ found in Eq. (\ref{e:stokes-drop}) for the
regime of Stokes flow and high dilution we obtain for the hindered
settling function,
\begin{equation}
  \label{e:hindered_settling}
  f_{\rm hs}(\Phi) = ( 1-\Phi )^3.
\end{equation}
For the description of experimental data one often uses the
phenomenological Richardson-Zaki correlation~\cite{Richardson54},
\begin{equation}
  \label{e:richardson-zaki}
  f_{\rm hs} (\Phi) = ( 1-\Phi )^n,
\end{equation}
where $n$ takes values around $5.$ Theoretical arguments by Batchelor
\cite{Batchelor72,Feuillebois84} lead to the expression
\begin{equation}
  \label{e:batchelor}
   f_{\rm hs}(\Phi) = 1- 6.55 \Phi
\end{equation}
for small $\Phi$.

For comparison with the simulation results we have plotted relation
(\ref{e:hindered_settling}) in Fig. \ref{f:sed-vel} as solid line.

\section{Conclusion and Outlook}
\label{s:conclusion}

\subsection{General advantages of a drag-force based approach}

The major advantage of a drag-force based algorithm as described here
for the simulation of two-phase flows is the capability to simulate
considerably large systems with relatively moderate computational
requirements. At the same time one gets without efforts a physical
behavior of the particulate phase circumventing the problems of
continuum approaches with stress or the constitutive equation of the
particulate phase. Such an algorithm is therefore a tool to obtain
quick insight into collective effects of many particles. In fact, most
of the known physical effects appear to be reproducible using only a
very simple expression for the momentum exchange between phases.

\subsection{Specific problems of a drag-force based approach}
\label{s:-problems}

We list now some of the problems that we have encountered performing
the present study that will be addressed in future research.

Let us first consider the behavior of a single particle in an
otherwise resting liquid, both initially at rest. If the particle
density is higher then the density of the liquid, the particle will
experience a net force in direction of gravity. As it falls, liquid
will be dragged down with it in accordance with the momentum exchange
rules and the requirements of the replacement of liquid as the local
liquid fraction changes. As compared to the liquid velocity far
away from the particle the local velocity is closer to the particle
velocity. Consequently the isolated particle experiences a drag force
which does not agree with the theoretical Stokes expression into which
the particle velocity and the liquid velocity at infinity enter and
the resulting terminal fall velocity of the particle is larger than
the theoretical Stokes velocity.\footnote{
  This effect has been taken into account by using the measured
  terminal velocity of a single sphere in lieu of the theoretical
  Stokes velocity, for example in the scaling of the mean
  sedimentation velocities to obtain the hindered settling function.  }

One way to address this problem is to alter the drag law
(\ref{e:stokes}) by introducing a ``drag-coefficient'' different
from $6\pi$ and dependent on the, albeit unphysical, ratio of particle
size to the grid-spacing and physical parameters as the Reynolds
number --- this approach has been taken in \cite{Kalthoff95}.
However, the root of this problem lies in the treatment of the momentum
exchange and indicates that a revision of its treatment is necessary.

One could imagine first that a better treatment of the momentum
exchange should proceed along the lines of approximating the liquid
stress tensor on the particle surface, which may be possible in the low
Reynolds number regime.  However, the principal limitation of our
{\it ansatz\/} is that the flow velocities in the vicinity of the
particle surface are not accurately rendered as required for an
estimation of the stress on the particle, since we account for
the flow only in an average sense.

Another possibility is to change the drag law. We have so far employed
a simple Stokes expression. It is probably necessary to introduce a
drag expression which additionally depends on the local particle
density in addition to the local velocity difference between phases.
We think that at least at high particle densities also particle
rotation becomes important and must be included into the model.
Moreover, we may have to use density dependent local viscosities in
the ``Navier-Stokes'' equation.

In conclusion, we should and will undertake an improvement of our
algorithm to tackle some of the problems listed above in the future.
However, even at the present state, we have a powerful tool at hand to
assess cooperative effects of many-particle systems in which
hydrodynamic interactions play an important role. We have successfully
modeled flow around impurities at small and moderate volume fractions
and found reasonable agreement in the case of a sedimenting system.

\section*{Acknowledgments}

We would like to acknowledge in particular long discussions with
Fran\c{c}ois Feuillebois, Hans Herrmann, Wolfgang Kalthoff, and Frank
Tzschichholz. We are also grateful for discussions with
D. Barnea, M. Fermigier, I. Goldhirsch, E. Guazzelli, and J. Hinch.
Stefan Schwarzer thanks the scientific council of the NATO
for financial support (granted through the DAAD, Bonn).


\newpage

\begin{table}[h]
\begin{tabular}{clrl}
 $L_x$          & system width            & $3$     & $cm$ \\
 $L_y$          & system height           & $6$     & $cm$ \\
 $z$            & Stokes force reference length & $0.03$ & $cm$ \\
 $\rho_{\rm l}$ & liquid density          & $1.09$  & $g/cm^3$  \\
 $\eta$         & liquid shear viscosity  & $0.4$ & $g/cm\,s$ \\
 $\bar{r}$      & average particle radius & $0.015$ & $cm$ \\
 $p$            & polydispersity          & $0.1$ & $1$ \\
\end{tabular}
\medskip
\caption{List of parameters for the simulation of flow through
  porous media. Deviating parameters are stated in the text.}
\label{t:flow}
\end{table}
\bigskip

\begin{table}[h]
\begin{tabular}{clrl}
 $L_x$          & system width            & $2.5$     & $cm$ \\
 $L_y$          & system height           & $5.0$     & $cm$ \\
 $z$            & Stokes force reference length & $0.05$ & $cm$ \\
 $\rho_{\rm l}$ & liquid density          & $1.0$  & $g/cm^3$  \\
 $\eta$         & liquid shear viscosity  & $0.4$ & $g/cm\,s$ \\
 $\rho_{\rm p}$ & particle density        & $2.53$  & $g/cm^3$  \\
 $\bar{r}$      & average particle radius & $0.025$ & $cm$ \\
 $p$            & polydispersity          & $0.01$ & $1$\\
 $k_n$          & repulsion constant      & $2.5\times 10^4$ & $g/s^2$ \\
 $\gamma_t$     & dissipation constant    & 80 & $1/s$ \\
 $\mu$          & Coulomb friction factor & 0.3 & $1$\\
 $g$            & gravitational acceleration & $981$ & $cm/s^2$ \\
\end{tabular}
\medskip
\caption{List of parameters for the simulation of batch settling
  sedimentation. Deviating parameters are stated in the text.}
\label{t:sed}
\end{table}

\newpage
\widetext

\bigskip

\newpage

\end{document}